\documentclass[12pt,oneside,reqno]{article}

\usepackage{amsmath}
\usepackage{amsthm,amsfonts}
\usepackage{amssymb,latexsym}
\usepackage{graphicx,rotating}
\usepackage{exscale} 
\usepackage{upref} 
\usepackage[latin1]{inputenc}   
\usepackage{txfonts} 
\usepackage{hyperref}

\theoremstyle{plain}

\theoremstyle{definition}

\theoremstyle{remark}

\let\mathbb=\varmathbb

\allowdisplaybreaks[1] 

\def\E{{\mathbb E}}

\begin{document}

\title{A Generalization of the Exponential-Poisson Distribution}

\author{Wagner Barreto-Souza and Francisco Cribari-Neto\\\\
Departamento de Estat\'{\i}stica\\
         Universidade Federal de Pernambuco\\
     Cidade Universit\'{a}ria\\
         Recife/PE, 50740--540, Brazil\\
e-mail: wagnerbs85@hotmail.com; cribari@de.ufpe.br}

\date{}
\maketitle

\begin{abstract}
The two-parameter distribution known as exponential-Poisson (EP) distribution, which has decreasing failure rate, was introduced by Kus (2007). In this paper we generalize the EP distribution and show that the failure rate of the new distribution can be decreasing or increasing. The failure rate can also be upside-down bathtub shaped. A comprehensive mathematical treatment of the new distribution is provided. We provide closed-form expressions for the density, cumulative distribution, survival and failure rate functions; we also obtain the density of the $i$th order statistic. We derive the $r$th raw moment of the new distribution and also the moments of order statistics. Moreover, we discuss estimation by maximum likelihood and obtain an expression for Fisher's information matrix. Furthermore, expressions for the Rényi and Shannon entropies are given and estimation of the stress-strength parameter is discussed. Applications using two real data sets are presented. 
\end{abstract}

{ \bf keywords}: Exponential-Poisson distribution, Generalized exponential-Poisson distribution, Failure rate, Order statistics, Fisher's information matrix.\\

\section{Introduction}\label{S:intro}
Adamidis and Loukas (1998) introduced a distribution with decreasing failure rate. This distribution is known as exponential-geometric distribution and is obtained by compouding an exponential distribution with a geometric distribution. In the same fashion, Kus (2007) introduced a two-parameter distribution known as exponential-Poisson (EP) distribution, which has decreasing failure rate, by compouding an exponential distribution with a Poisson distribution. Proschan (1963) proved that the decreasing failure rate property is inherent to mixtures of distributions with constant failure rates. 
The cumulative distribution function (cdf) of the EP distribution is given by
\begin{equation*}
G_{EP}(x)=\frac{1-e^{-\lambda+\lambda\exp(-\beta x)}}{1-e^{-\lambda}},
\end{equation*}
$x,\beta,\lambda>0$.

Gupta and Kundu (1999) introduced the generalized exponential (GE) distribution (also known as exponentiated exponential distribution) whose the cdf is 
\begin{equation*}
G_{GE}(x)=(1-e^{-\lambda x})^\alpha, \quad x,\lambda,\alpha>0.
\end{equation*}

Following their approach, in this paper we generalize the EP distribution by considering the following cdf: 
\begin{equation}\label{cdf}
F(x)=G_{EP}(x)^\alpha=\left(\frac{1-e^{-\lambda+\lambda\exp(-\beta x)}}{1-e^{-\lambda}}\right)^\alpha,
\end{equation}
$\alpha>0$. A random variable $X$ with cdf (\ref{cdf}) is said to follow a generalized exponential-Poisson (GEP) distribution (or exponentiated exponential-Poisson distribution).

Other distributions obtained using this approach have been introduced and studied in the literature. The exponentiated Weibull (EW) distribution introduced by Mudholkar et al.\ (1993) generalizes the GE distribution and was also considered by Mudholkar et al.\ (1995), Mudholkar and Hutson (1996) and Nassar and Eissa (2003). Generalizations of the gamma, Weibull, Gumbel and Fréchet distributions are given by exponentiated type distributions: the exponentiated gamma, exponentiated Weibull, exponentiated Gumbel and exponentiated Fréchet distributions, respectively; see Nadarajah and Kotz (2006).

Generalizations of exponentiated type distributions can be obtained from the class of generalized beta distributions. This class has been receiving increased attention recently, in particular after the works of Eugene et al.\ (2002) and Jones (2004). Eugene et al.\ (2002) introduced what is known as the beta normal (BN) distribution; the exponentiated normal distribution is a particular case of the BN distribution. Nararajah and Kotz (2004) introduced the beta Gumbel distribution which generalizes the exponentiated Gumbel distribution. Nadarajah and Kotz (2005) also introduced the beta exponential (BE) distribution which has the GE distribution as a particular case. A generalization of the exponentiated Fréchet distribution was obtained by Nadarajah and Gupta (2004): the beta Fréchet distribution.

The remainder of our paper is organized as follows. In Section \ref{S:model}, we give the density and failure rate functions of the GEP distribution. Furthermore, we show that the failure rate can be decreasing or decreasing; it can also be upside-down bathtub. Additionally, we derive the probability density function of the $i$th order statistic. Expressions for the $r$th raw moments of the GEP distribution and of the $i$th order statistic are given in Section \ref{S:quantilesmoments}. We discuss estimation by maximum likelihood and provide an expression for Fisher's information matrix in Section \ref{S:inference}. In Section \ref{S:entropy} we give expressions for the Rényi and Shannon entropies. We discuss estimation of the stress-strength parameter in Section \ref{S:SSR}. Applications are given in the Section \ref{S:applications}. Finally, Section \ref{S:conclusions} concludes the paper.

\section{Density, failure rate functions and order statistic}\label{S:model}

Let $X$ be a random variable with GEP distribution with parameters $\lambda$, $\beta$ and $\alpha$, i.e., $X\sim GEP(\lambda,\beta,\alpha)$. The cdf of $X$ is given in (\ref{cdf}) and the associated probability density function (pdf) is 
\begin{equation}\label{pdf}
f(x)=\frac{\alpha\lambda\beta}{(1-e^{-\lambda})^\alpha}\{1-e^{-\lambda+\lambda\exp(-\beta x)}\}^{\alpha-1}e^{-\lambda-\beta x+\lambda\exp(-\beta x)},
\end{equation}
$x,\lambda,\beta,\alpha>0$.

The EP distribution is a particular case of the GEP distribution corresponding $\alpha=1$. When $\alpha=1$ and taking $\lambda\rightarrow0^{+}$, it can be shown that $X \leadsto Y$, where `$\leadsto$' denotes convergence in distribution and $Y$ is a random variable with exponential distribution with parameter $\beta$. 

\begin{figure}[ht!]
	\centering
		\includegraphics[width=0.6\textwidth]{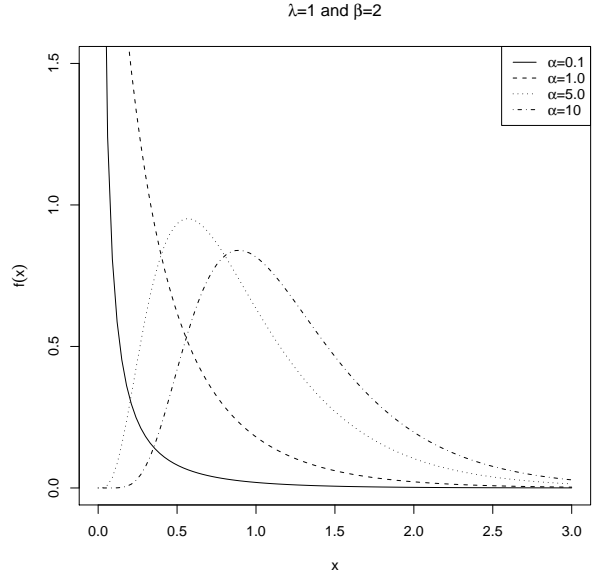}
	\caption{GEP densities.}
	\label{fig:figura1}
\end{figure}

For $|z|<1$ and $\gamma>0$, we have that 
\begin{equation}\label{expansion}
(1-z)^{\gamma-1}=\sum_{j=0}^\infty\frac{(-1)^j\Gamma(\gamma)}{\Gamma(\gamma-j)j!}z^j.
\end{equation}
If $\gamma$ is integer, the sum in (\ref{expansion}) stops at $\gamma-1$.
Let $\alpha>0$ be non-integer. Plugging (\ref{expansion}) into (\ref{pdf}), we obtain
\begin{equation}\label{pdfexpreal}
f(x)=\frac{\Gamma(\alpha+1)}{(1-e^{-\lambda})^\alpha}\sum_{j=0}^\infty \frac{(-1)^j\{1-e^{-\lambda(j+1)}\}}{\Gamma(\alpha-j)(j+1)!}f_{EP}(x;\lambda(j+1),\beta), 
\end{equation}
where $f_{EP}(\cdot,\lambda(j+1),\beta)$ is the pdf of a random variable with EP distribution with parameters $\lambda(j+1)$ and $\beta$. When $\alpha>0$ is integer, the index $j$ in the summation in (\ref{pdfexpreal}) ranges up to $\alpha-1$. We then have that the the GEP pdf can be written as a linear combination of EP pdfs. Figure \ref{fig:figura1} shows different GEP densities, i.e., it contains plots of (\ref{pdf}).

The survival function and the failure rate of the GEP distribution are given by 
\begin{equation*}
s(x)=1-\left(\frac{1-e^{-\lambda+\lambda\exp(-\beta x)}}{1-e^{-\lambda}}\right)^\alpha, \quad x>0, 
\end{equation*}
and
\begin{equation}\label{failure}
h(x)=\alpha\lambda\beta\frac{\{1-e^{-\lambda+\lambda\exp(-\beta x)}\}^{\alpha-1}e^{-\lambda-\beta x+\lambda\exp(-\beta x)}}{(1-e^{-\lambda})^\alpha-\{1-e^{-\lambda+\lambda\exp(-\beta x)}\}}, \quad x>0,
\end{equation}
respectively.

We shall now show that the failure rate of the GEP distribution can be decreasing or increasing depending on the parameter values. Define the function $\eta(x)=-f'(x)/f(x)$, where $f'$ denotes the first derivative of $f$. It is straightfoward to show that  
$$\eta(x)=-\lambda\beta(\alpha-1)e^{-\lambda}\{1-e^{-\lambda+\lambda\exp(-\beta x)}\}^{-1}e^{-\beta x +\lambda\exp(-\beta x)}+\beta(\lambda e^{-\beta x}+1)$$
and 
\begin{align*}
\eta'(x)&=\lambda^2\beta^2(\alpha-1)e^{-2\lambda}\{1-e^{-\lambda+\lambda\exp(-\beta x)}\}^{-2}e^{-2\beta x+2\lambda\exp(-\beta x)}+\lambda\beta^2(\alpha-1)e^{-\lambda}\\
&{}\times\{1-e^{-\lambda+\lambda\exp(-\beta x)}\}^{-1}\{1+\lambda e^{-\beta x}\}e^{-\beta x+\lambda\exp(-\beta x)}-\lambda\beta^2e^{-\beta x}.
\end{align*}
If $0<\alpha\leqslant 1$, then $\eta'(x)<0\,\, \forall x>0$. It follows from Theorem (b) of Glaser (1980) that the failure rate is decreasing.   

\begin{figure}[ht!]
	\centering
		\includegraphics[width=0.5\textwidth]{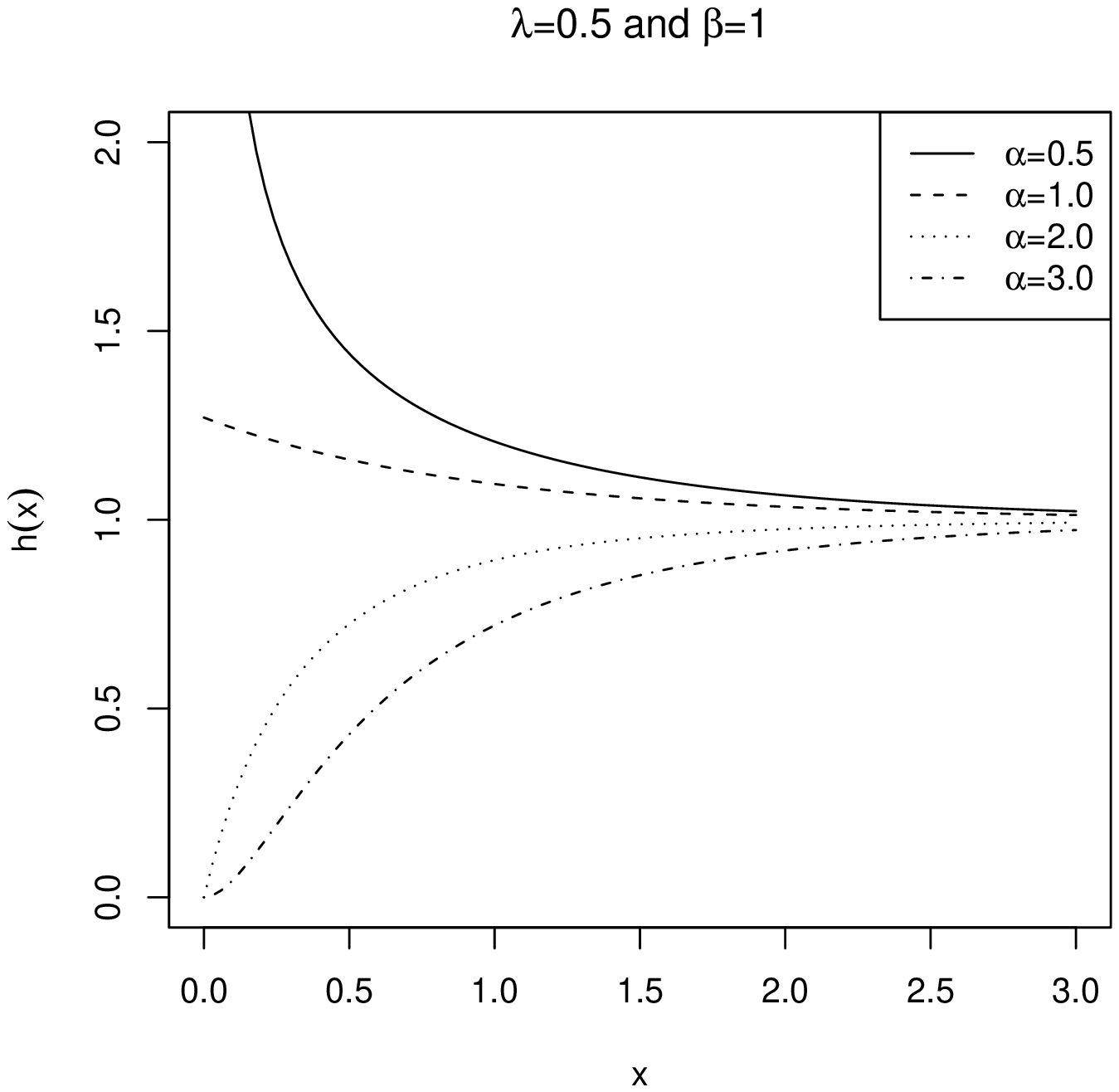}\includegraphics[width=0.5\textwidth]{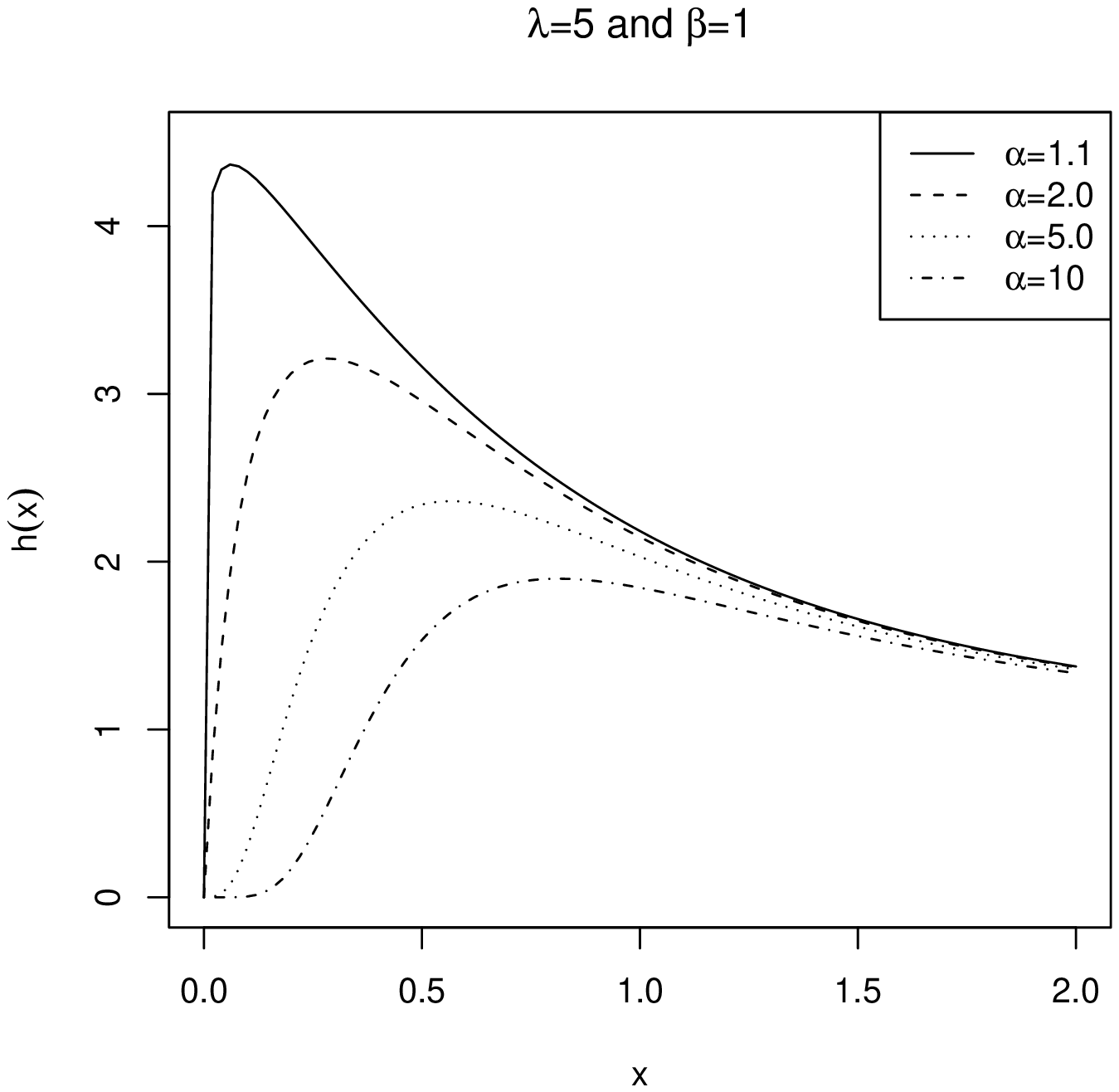}
	\caption{GEP failure rates.}
	\label{fig:figura2}
\end{figure}

Using the inequalities 
\begin{equation*}
\frac{e^{-\lambda+\lambda\exp(-\beta x)}}{1-e^{-\lambda+\lambda\exp(-\beta x)}}\geqslant \frac{1}{e^\lambda-1} \quad \mbox{and} \quad 1<1+\lambda e^{-\beta x}\leqslant 1+\lambda,
\end{equation*}
we obtain 
$$\eta'(x)\geqslant \lambda\beta^2e^{-\beta x}\left\{\lambda(\alpha-1)\frac{e^{-\beta x}}{(e^\lambda-1)^2}+\frac{\alpha-1}{e^\lambda-1}-1 \right\}.$$
Therefore, if $\alpha>e^\lambda$, then $\eta'(x)>0\,\, \forall x>0$. Hence, using Theorem (b) of Glaser (1980), we conclude that the failure rate is increasing. Figure \ref{fig:figura2} shows plots of the failure rate given in (\ref{failure}). We note that, as mentioned earlier, these rates can be increasing or decreasing. Furthermore, we see that failure rate can be upside-down bathtub. 

The pdf $f_{i:n}$ of the $i$th order statistic for a random sample $X_1,\ldots,X_n$ from the GEP distribution is given by
\begin{equation*}
f_{i:n}(x)=\frac{n!}{(i-1)!(n-i)!}f(x)F(x)^{i-1}\{1-F(x)\}^{n-i}, 
\end{equation*}
and then 
\begin{align*}
f_{i:n}(x)&=\frac{n!}{(i-1)!(n-i)!}\frac{\alpha\lambda\beta}{(1-e^{-\lambda})^{\alpha n}}\{1-e^{-\lambda+\lambda\exp(-\beta x)}\}^{\alpha i-1}\nonumber\\
&{}\times e^{-\lambda-\beta x+\lambda\exp(-\beta x)}\{(1-e^{-\lambda})^\alpha-e^{-\lambda+\lambda\exp(-\beta x)}\}^{\alpha(n-i)}, 
\end{align*}
where $f(\cdot)$ and $F(\cdot)$ are pdf and cdf of the GEP distribution, respectively.

An useful alternative expression for the pdf of the $i$th order statistc is
\begin{equation}\label{pdforder}
f_{i:n}(x)=\frac{n!}{(i-1)!(n-i)!}\sum_{k=0}^{n-i}\frac{(-1)^k}{i+k}\binom{n-i}{k}f_{GEP}(x;\lambda,\beta,\alpha(i+k)),
\end{equation}
where $f_{GEP}(.;\lambda,\beta,\alpha(i+j))$ is the pdf of a random variable having GEP distribution with parameters $\lambda$, $\beta$ and $\alpha(i+j)$. The pdf of the $i$th order statistic for a random sample of the EP distribution is obtained by setting $\alpha=1$ in (\ref{pdforder}). Expressions (\ref{pdfexpreal}) and (\ref{pdforder}) play an important role in the derivation of the main properties of the GEP distribution. 

\section{Quantiles and moments}\label{S:quantilesmoments}

In this section we shall present closed-form expressions for the GEP quantiles and also for the raw moments of the distribution order statistics. 
The $q$th quantile $x_q$ of the GEP distribution can be obtained from (\ref{cdf}) as
\begin{equation*}
x_q=-\beta^{-1}\log\left[ 1+\lambda^{-1}\log\{1-q^{1/\alpha}(1-e^{-\lambda})\}\right].
\end{equation*}
In particular, the distribution median is 
\begin{equation*}
x_{0.5}=-\beta^{-1}\log\left[ 1+\lambda^{-1}\log\{1-2^{-1/\alpha}(1-e^{-\lambda})\}\right]. 
\end{equation*}

We shall now provide a closed-form expression for the $r$th raw moment of the GEP distribution. Kus (2007) showed that the $r$th moment of a random variable $Y$ with EP distribution with parameters $\lambda$ and $\beta$ can be expressed as
\begin{equation}\label{momentep}
\E(Y^r)=\frac{r!\lambda }{(e^\lambda-1)\beta^r}F_{r+1,r+1}([1,\ldots,1],[2,\ldots,2],\lambda),
\end{equation}
where $$F_{p,q}({\bf n},{\bf d})=\sum_{k=0}^\infty\frac{\lambda^k\prod_{i=1}^p\Gamma(n_i+k)\Gamma(n_i)^{-1}}{k!\prod_{i=1}^q\Gamma(d_i+k)\Gamma(d_i)^{-1}},$$
with ${\bf n}=[n_1,\ldots,n_p]$ and ${\bf d}=[d_1,\ldots,d_q] $. The latter is the known Barnes extended hypergeometric function, which can be easily evaluated using, e.g., the computer algebra systems \textsc{Maple} and \textsc{Mathematica}. 

Let $X\sim GEP(\lambda,\beta,\alpha)$. If $\alpha>0$ is non-integer, using (\ref{pdfexpreal}) and (\ref{momentep}) we obtain that the $r$th raw moment of $X$ can be written as 
\begin{equation}\label{momentreal}
\E(X^r)=\frac{r!\Gamma(\alpha+1)\lambda}{(1-e^{-\lambda})^{\alpha}\beta^r}\sum_{j=0}^\infty\frac{(-1)^je^{-\lambda(j+1)}}{\Gamma(\alpha-j)j!}F_{r+1,r+1}([1,\ldots,1],[2,\ldots,2],\lambda(j+1)); 
\end{equation}
when $\alpha>0$ is integer the above summation stops at $\alpha-1$.

Let $X_{i:n}$ be the $i$th order statistic of a random sample from the GEP distribution. From (\ref{pdforder}), we have that the $r$th raw moment of $X_{i:n}$ for $\alpha>0$ non-integer can be written as 
\begin{align}\label{momentorderreal}
\E(X_{i:n}^r)&=\frac{n!r!\lambda}{(i-1)!(n-i)!\beta^r}\sum_{j=0}^\infty\sum_{k=0}^{n-i}\frac{(-1)^{j+k}\Gamma(\alpha(i+k)+1)e^{-\lambda(j+1)}}{\Gamma(\alpha(i+k)-j)j!(i+k)(1-e^{-\lambda})^{\alpha(i+k)}}\nonumber\\
&{}\times \binom{n-i}{k}F_{r+1,r+1}([1,\ldots,1],[2,\ldots,2],\lambda(j+1)); 
\end{align}
when $\alpha>0$ is integer the index $j$ in the above summation only ranges up tp $\alpha-1$. It is noteworthy that one can obtain the moments of the order statistics of the EP distribution by setting $\alpha=1$ in (\ref{momentorderreal}), a result which was not, to the best of our knowledge, available in the literature. That is, the above result yields as a by product the raw EP order statistics moments. One can also use such a result to compute raw moments of order statistics of a random GEP sample.

\section{Estimation and testing}\label{S:inference}

In what follows we shall discuss point and interval estimation, and also hypothesis testing inference on the parameters that index the GEP distribution. 
Let $\theta=(\lambda,\beta,\alpha)^\top$ be the parameter vector. The log-likelihood $\ell=\ell(\theta)$ for a single observation $x$ of the random variable $X\sim GEP(\lambda,\beta,\alpha)$ is 
\begin{align}\label{loglik}
\ell&=-\lambda+\log(\alpha\lambda\beta)-\alpha\log(1-e^{-\lambda})
+(\alpha-1)\log\{1\nonumber\\
&-e^{-\lambda+\lambda\exp(-\beta x)}\} -\beta x+\lambda e^{-\beta x},\quad x>0.
\end{align} 
The score function for (\ref{loglik}) is given by $U(\theta)=(\partial \ell/\partial \lambda,\partial \ell/\partial \beta,\partial \ell/\partial \alpha)^\top$, where
\begin{align*}
\frac{\partial\ell}{\partial\lambda}&=\frac{1}{\lambda}-\frac{\alpha}{e^\lambda-1}+(e^{-\beta x}-1)[1-(1-\alpha)\{e^{\lambda-\lambda\exp(-\beta x)}-1\}^{-1}],\\
\frac{\partial\ell}{\partial\beta}&=\frac{1}{\beta}-x+\lambda xe^{-\beta x}[(\alpha-1)\{e^{\lambda-\lambda\exp(-\beta x)}-1\}^{-1}-1],\\
\frac{\partial\ell}{\partial\alpha}&=\frac{1}{\alpha}-\log(1-e^{-\lambda})+\log\{1-e^{-\lambda+\lambda\exp(-\beta x)}\}.
\end{align*}
As is well known, the expected value of the score function equals zero, i.e., $\E(U(\theta))=0$, which implies 
\begin{align}\label{eq1}
\E\left[\log\{1-e^{-\lambda+\lambda\exp(-\beta X)}\}\right] &= \log(1-e^{-\lambda})-\frac{1}{\alpha},\\
\E\left[(e^{-\beta X}-1)\{1-(\alpha-1)(e^{\lambda-\lambda\exp(-\beta X)}-1)^{-1}\}\right]&= \frac{\alpha}{e^\lambda-1}-\frac{1}{\lambda}+1,\nonumber\\
\E\left[e^{-\beta X}X\{(\alpha-1)(e^{\lambda-\lambda\exp(-\beta X)}-1)^{-1}-1\}\right]&= \lambda^{-1}(\E(X)-\beta)\nonumber.
\end{align}

The total log-likelihood of the random sample $x=(x_1,\ldots,x_n)^\top$ of size $n$ from $X$ is given by $\ell_n=\sum_{i=1}^n\ell^{(i)}$ and the total score function is given by $U_n=\sum_{i=1}^nU^{(i)}$, where $\ell^{(i)}$ is the log-likelihood of the $i$th observation and $U^{(i)}$ is as given above, for $i=1,\ldots,n$. The maximum likelihood estimator $\hat{\theta}$ of $\theta$ is obtained by numerically solving the nonlinear system of equations $U_n=0$. It is usually more convenient, however, to use a nonlinear optimization algorithm (such as the quasi-Newton algorithm known as BFGS) to numerically maximize the log-likelihood function in (\ref{loglik}).  

Fisher's information matrix is given by
\begin{align*}\label{fisher}
K = K_n(\theta) = n\left[ \begin{array}{ccc}
\kappa_{\lambda,\lambda}&\kappa_{\lambda,\beta}&\kappa_{\lambda,\alpha}\\
\kappa_{\lambda,\beta}&\kappa_{\beta,\beta}&\kappa_{\beta,\alpha}\\
\kappa_{\lambda,\alpha}&\kappa_{\beta,\alpha}&\kappa_{\alpha,\alpha}\\
\end{array}\right],
\end{align*}
where
\begin{align*}
\kappa_{\lambda,\lambda}&=\frac{1}{\lambda^2}-\frac{\alpha e^\lambda}{(e^\lambda-1)^2}+(\alpha-1)e^\lambda \E\left[ e^{-\lambda\exp(-\beta X)}\left( \frac{e^{-\beta X}-1}{e^{\lambda-\lambda\exp(-\beta X)}-1}\right)^2\right],\\
\kappa_{\lambda,\beta}&=-\frac{1}{\lambda}\left\{\E(X)-\frac{1}{\beta}\right\}-\lambda(\alpha-1)e^\lambda \E\left[X\frac{e^{-\beta X}(e^{-\beta X}-1)e^{-\lambda\exp(-\beta X)}}{(e^{\lambda-\lambda\exp(-\beta X)}-1)^2}\right],\\
 \kappa_{\alpha,\lambda}&=\frac{1}{\alpha-1}\left\{1+\frac{\alpha}{e^\lambda-1}-\frac{1}{\lambda}-\E(e^{-\beta X})\right\}, \quad \kappa_{\alpha,\alpha}=\frac{1}{\alpha^2},\\
 \kappa_{\beta,\beta}&=\frac{1}{\beta^2}+\lambda \E\left[X^2e^{-\beta X}\{(\alpha-1)(e^{\lambda-\lambda\exp(-\beta X)}-1)^{-1}-1\} \right]\\
  &{}+ (\alpha-1)\lambda^2e^\lambda \E\left[X^2\frac{e^{-2\beta X}e^{-\lambda\exp(-\beta X)}}{(e^{\lambda-\lambda\exp(-\beta X)}-1)^2} \right],\\
  \kappa_{\beta,\alpha}&=\frac{1}{\alpha-1}\left\{\frac{1}{\beta}-\E(X)-\lambda \E(Xe^{-\beta X})\right\}.
\end{align*}
The above expressions depend on some expectations that can be easily solved computed using numerical integration.

Under the usual regularity conditions, the asymptotic distribution of
$$\sqrt n (\hat\theta-\theta)\,\,\,\,\mathrm{is}\,\,\,\,\mathcal{N}_3(0,K(\theta)^{-1}),$$ where
$\lim_{n\rightarrow\infty}K_n(\theta)^{-1}=K(\theta)^{-1}$. It is noteworthy that the multivariate normal distribution $\mathcal{N}_3(0,K_n(\hat\theta)^{-1})$ can be used to construct confidence intervals for the parameters. In fact, an $(1-\gamma)\times 100\%$ ($0 < \gamma < 1/2$) asymptotic confidence interval for the $i$th parameter $\theta_i$ in $\theta$ is 
$$ACI_i=(\hat\theta_i-z_{1-\gamma/2}\sqrt{\hat\kappa^{\theta_i,\theta_i}},\hat{\theta_i}
+z_{1-\gamma/2}\sqrt{\hat\kappa^{\theta_i,\theta_i}}),$$
where $\hat\kappa^{\theta_i,\theta_i}$ denotes the $i$th diagonal element of
$K_n(\hat\theta)^{-1}$ for $i=1,2,3$ and $z_{1-\gamma/2}$ is the $1-\gamma/2$ standard normal quantile. 

We shall now move to hypothesis testing inference on the parameters that index de GEP law. Consider the partition $\theta=(\theta_1^\top,\theta_2^\top)^\top$ of the GEP parameter vector and suppose we wish to test the hypothesis $\mathcal{H}_0:\theta_1 =\theta_1^{(0)}$ against the alternative hypothesis $\mathcal{H}_A:\theta_1 \neq \theta_1^{(0)}$. To that end, we can use the likelihood ratio ($LR$) test whose test statistic is given by 
$w= 2\{\ell(\hat{\theta})-\ell(\tilde{\theta})\}$, where
$\hat\theta = (\hat\theta_1^\top, \hat\theta_2^\top)^\top$ and $\tilde\theta = ((\theta_1^{(0)})^\top, \tilde\theta_2^\top)^\top$ denote the MLEs of $\theta$ under the null and the alternative hypotheses, respectively. Under the null hypothesis, $w$ is asymptotically (as $n\to\infty$) distributed as $\chi_k^2$, where $k$ is the dimension of the vector $\theta_1$ of parameters of interest. We reject the null hypothesis at the nominal level $\gamma$ ($0 < \gamma < 1$) if $w > \chi^2_{k, 1-\gamma}$, where $\chi^2_{k, 1-\gamma}$ is the $1-\gamma$ quantile of $\chi_k^2$. Using this test, one can select between a GEP and an EP model, which can be done by testing $\mathcal{H}_0:\alpha=1$ against $\mathcal{H}_A:\alpha\ne 1$.

\section{Rényi and Shannon entropies}\label{S:entropy}   

The entropy of a random variable $X$ is a measure of uncertainty 
variation. The Rényi entropy is defined as $I_R(\gamma)=\frac{1}{1-\gamma}\log\{\int_{\mathbb{R}} f^\gamma(x)dx\}$ where $\gamma>0$ and $\gamma\neq1$. 
We have that
\begin{equation*}
\int_0^\infty f^\gamma(x)dx=\left\{\frac{\alpha\lambda\beta e^{-\lambda}}{(1-e^{-\lambda})^\alpha}\right\}^\gamma\int_0^\infty\{1-e^{-\lambda+\lambda\exp(-\beta x)}\}^{(\alpha-1)\gamma}e^{-\gamma\beta x+\gamma\lambda\exp(-\beta x)}dx.
\end{equation*}

For $(\alpha-1)\gamma+1>0$, we used (\ref{expansion}) to expand the term $\{1-e^{-\lambda+\lambda\exp(-\beta x)}\}^{(\alpha-1)\gamma}$ in the previous integral and obtain
\begin{equation*}
\int_0^\infty f^\gamma(x)dx=\left\{\frac{\alpha\lambda\beta e^{-\lambda}}{(1-e^{-\lambda})^\alpha}\right\}^\gamma\sum_{j=0}^\infty \frac{(-1)^j\Gamma((\alpha-1)\gamma+1)}{\Gamma((\alpha-1)\gamma+1-j)j!\beta}A_j(\lambda,\gamma),
\end{equation*}
where $A_j(\lambda,\gamma)=\int_0^\infty e^{-\gamma\beta x+\lambda(\gamma+j)\exp(-\beta x)}dx=\sum_{k=0}^\infty\frac{\{\lambda(\gamma+j)\}^k}{k!(k+\gamma)}$.\\

Hence, the Rényi entropy for GEP distribution can be expressed as 
\begin{equation*}
I_R(\gamma)=\frac{1}{1-\gamma}\log\left\{\beta^{\gamma-1} \frac{(\alpha\lambda e^{-\lambda})^\gamma}{(1-e^{-\lambda})^{\alpha\gamma}}\sum_{j=0}^\infty \frac{(-1)^j\Gamma((\alpha-1)\gamma+1)}{\Gamma((\alpha-1)\gamma+1-j)j!}A_j(\lambda,\gamma)\right\}.
\end{equation*}

Shannon's entropy is defined as $\E[-\log f(X)]$ and can be obtained by taking $\lim_{\gamma\uparrow1}I_R(\gamma)$. In our case, 
\begin{align*}
\E[-\log f(X)]&=\lambda-\log(\alpha\lambda\beta)+\alpha\log(1-e^{-\lambda})+\beta \E(X)-\lambda \E(e^{-\beta X})\\
& -(\alpha-1)\E(\log\{1-e^{-\lambda+\lambda\exp(-\beta x)}\}).
\end{align*}

After changing the variable $u=e^{-\lambda+\lambda\exp(-\beta x)}$ and using the expansion $\log x= -\sum_{k=1}^\infty \frac{(1-x)^k}{k}$ (valid for $0<x\leq1$), it becomes
\begin{align*}
\E(e^{-\beta X})&=\frac{\alpha\lambda\beta}{(1-e^{-\lambda})^\alpha}\int_0^\infty \{1-e^{-\lambda+\lambda\exp(-\beta x)}\}^{\alpha-1}e^{-\lambda-2\beta x+\lambda\exp(-\beta x)}dx\\
&=\frac{\lambda^{-1}\alpha}{(1-e^{-\lambda})^\alpha}\int_{e^{-\lambda}}^1 \log u(1-u)^{\alpha-1}du+1
=\frac{\alpha}{\lambda}\sum_{k=1}^\infty\frac{(1-e^{-\lambda})^k}{k(k+\alpha)}.
\end{align*}

Finally, using (\ref{momentreal}) and (\ref{eq1}) we obtain
\begin{align*}
\E[-\log f(X)]&=-\log(\alpha\lambda\beta)+\log(1-e^{-\lambda})+\frac{\alpha-1}{\alpha}-\alpha\sum_{k=1}^\infty\frac{(1-e^{-\lambda})^k}{k(k+\alpha)}\\
&+\frac{\lambda\Gamma(\alpha+1)}{(1-e^{-\lambda})^\alpha}\sum_{k=0}^\infty\frac{(-1)^ke^{-\lambda(k+1)}}{\Gamma(\alpha-k)k!}F_{2,2}([1,1],[2,2],\lambda(k+1)).
\end{align*} 

\section{Estimation of the stress-strength parameter}\label{S:SSR} 

We shall now discuss estimation of the stress-strength parameter for GEP distribtuion, which is defined as $R=\Pr(Y<X)$, where
$X\sim GEP(\lambda,\beta,\alpha_1)$ and $Y\sim GEP(\lambda,\beta,\alpha_2)$ are independent random variables.
Estimation of $R$ is oftentimes of interest in statistics and has received considerable attention in the last few decades. Estimation of the $\Pr(Y<X)$ when $X$ and $Y$ are normally distributed was considered by Govidarajulu (1967) and Church and Harris (1970). Constantine and Karson (1986) considered the estimation of $P(Y<X)$, when $X$ and $Y$ are independent gamma random variables. 
In our case, the stress-strength parameter $R$ is given by
\begin{equation*}
R=\frac{\alpha_1\lambda\beta e^{-\lambda}}{(1-e^{-\lambda})^{\alpha_1+\alpha_2}}
\int_0^\infty\{1-e^{-\lambda+\lambda\exp(-\beta x)}\}^{\alpha_1+\alpha_2-1}
e^{-\beta x+\lambda\exp(-\beta x)}dx=\frac{\alpha_1}{\alpha_1+\alpha_2}.
\end{equation*}
Let $X_1,\ldots,X_n$ and $Y_1,\ldots,Y_m$ denote independent random samples from $GEP(\lambda,\beta,\alpha_1)$ and $GEP(\lambda,\beta,\alpha_2)$ distributions, respectively.
The total log-likelihood $\ell_R(\theta^*)$ of the random samples $X_1,\ldots,X_n$ and $Y_1,\ldots,Y_m$ is given by
\begin{align}\label{loglikR}
\ell_R(\theta^*)&=(n+m)[\log(\lambda\beta)-\lambda]+n\log\alpha_1+m\log\alpha_2-(\alpha_1n+\alpha_2m)\nonumber\\
&\times \log(1-e^{-\lambda})+(\alpha_1-1)\sum_{i=1}^n\log\{1-e^{-\lambda+\lambda\exp(-\beta x_i)}\}+
(\alpha_2-1)\nonumber\\
&\times \sum_{i=1}^m\log\{1-e^{-\lambda+\lambda\exp(-\beta y_i)}\}+\lambda\left[\sum_{i=1}^ne^{-\beta x_i}+\sum_{i=1}^me^{-\beta y_i}\right].
\end{align}

The score function associated to the log-likelihood in (\ref{loglikR}) is \\
$U_R(\theta^*)=(\partial \ell_R/\partial \lambda,\partial \ell_R/\partial \beta, \partial \ell_R/\partial \alpha_1,\partial \ell_R/\partial \alpha_2)^\top$, where
\begin{align*}
\frac{\partial\ell_R}{\partial \lambda}&= -\frac{\alpha_1n+\alpha_2m}{e^\lambda-1}+\sum_{i=1}^n(1-e^{-\beta x_i})\{1+(\alpha_1-1)[e^{\lambda-\lambda\exp(-\beta x_i)}-1]^{-1}\}\\
&+\frac{n+m}{\lambda}+\sum_{i=1}^m(1-e^{-\beta y_i})\{1+(\alpha_2-1)[e^{\lambda-\lambda\exp(-\beta y_i)}-1]^{-1}\},\\
\frac{\partial\ell_R}{\partial \beta}&=\frac{n+m}{\beta}+\lambda\sum_{i=1}^nx_ie^{-\beta x_i} \left\{\frac{\alpha_1-1}{e^{\lambda-\lambda\exp(-\beta x_i)}-1}-1 \right\}\\
&+\lambda\sum_{i=1}^my_ie^{-\beta y_i} \left\{\frac{\alpha_2-1}{e^{\lambda-\lambda\exp(-\beta y_i)}-1}-1 \right\},\\
\frac{\partial\ell_R}{\partial \alpha_1}&=\frac{n}{\alpha_1}-n\log(1-e^{-\lambda})+\sum_{i=1}^n\log\{1-e^{-\lambda+\lambda\exp(-\beta x_i)}\},\\
\frac{\partial\ell_R}{\partial \alpha_2}&=\frac{m}{\alpha_2}-m\log(1-e^{-\lambda})+\sum_{i=1}^m\log\{1-e^{-\lambda+\lambda\exp(-\beta y_i)}\}.
\end{align*}

The maximum likelihood estimator of $\theta^*$ is $\hat{\theta}^*$, where $\hat{\theta}^*$ is obtained numerically by solving the nonlinear system of equations $U_R(\hat{\theta}^*)=0$. With this, we have that maximum likelihood estimator of $R$ is $\hat{R}=\hat{\alpha}_1/(\hat{\alpha}_1+\hat{\alpha}_2)$.

\section{Applications}\label{S:applications}

In this section we fit GEP distributions to two real data sets. The first data set is given by Hinkley (1977) and consists of thirty sucessive values of March precipitation (in inches) in Minneapolis/St Paul. The data are 0.77, 1.74, 0.81, 1.2, 1.95, 1.2, 0.47, 1.43, 3.37, 2.2, 3, 3.09, 1.51, 2.1, 0.52, 1.62, 
1.31, 0.32, 0.59, 0.81, 2.81, 1.87, 1.18, 1.35, 
4.75, 2.48, 0.96, 1.89, 0.9, 2.05.

The MLEs and the values of the maximized log-likelihoods for the EP and GEP distributions are 
\begin{equation*}
\hat{\lambda}=31.9785,\,\, \hat{\beta}=0.0186,\,\, \ell_{EP}(\hat{\theta})=-45.7935
\end{equation*}
and  
\begin{equation*}
\hat{\lambda}=0.8003,\,\, \hat{\beta}=0.7336,\,\, \hat{\alpha}=2.7329,\,\, \ell_{EP}(\hat{\theta})=-39.7229,
\end{equation*}
respectively.

The source of the second data set is The Open University (1993). The following data are the prices of the 31 different children's wooden toys on sale in a Suffolk craft shop in April 1991: 4.2, 1.12, 1.39, 2, 3.99, 2.15, 1.74, 5.81, 1.7, 2.85, 0.5, 0.99, 11.5, 5.12, 0.9, 1.99, 6.24, 2.6, 3, 12.2, 7.36, 4.75, 11.59, 8.69, 9.8, 1.85, 1.99, 1.35, 10, 0.65, 1.45.

The parameter MLEs and the maximized log-likelihoods for the EP model are 
\begin{equation*}
\hat{\lambda}=30.8795,\,\, \hat{\beta}=0.0077,\,\, \ell_{EP}(\hat{\theta})=-75.9447.
\end{equation*}
For the GEP model we obtain       
\begin{equation*}
\hat{\lambda}=1.9821,\,\, \hat{\beta}=0.2369,\,\, \hat{\alpha}=2.3144,\,\, \ell_{EP}(\hat{\theta})=-73.6629.
\end{equation*}

\begin{figure}[htbp]
	\centering
		\includegraphics[width=0.50\textwidth]{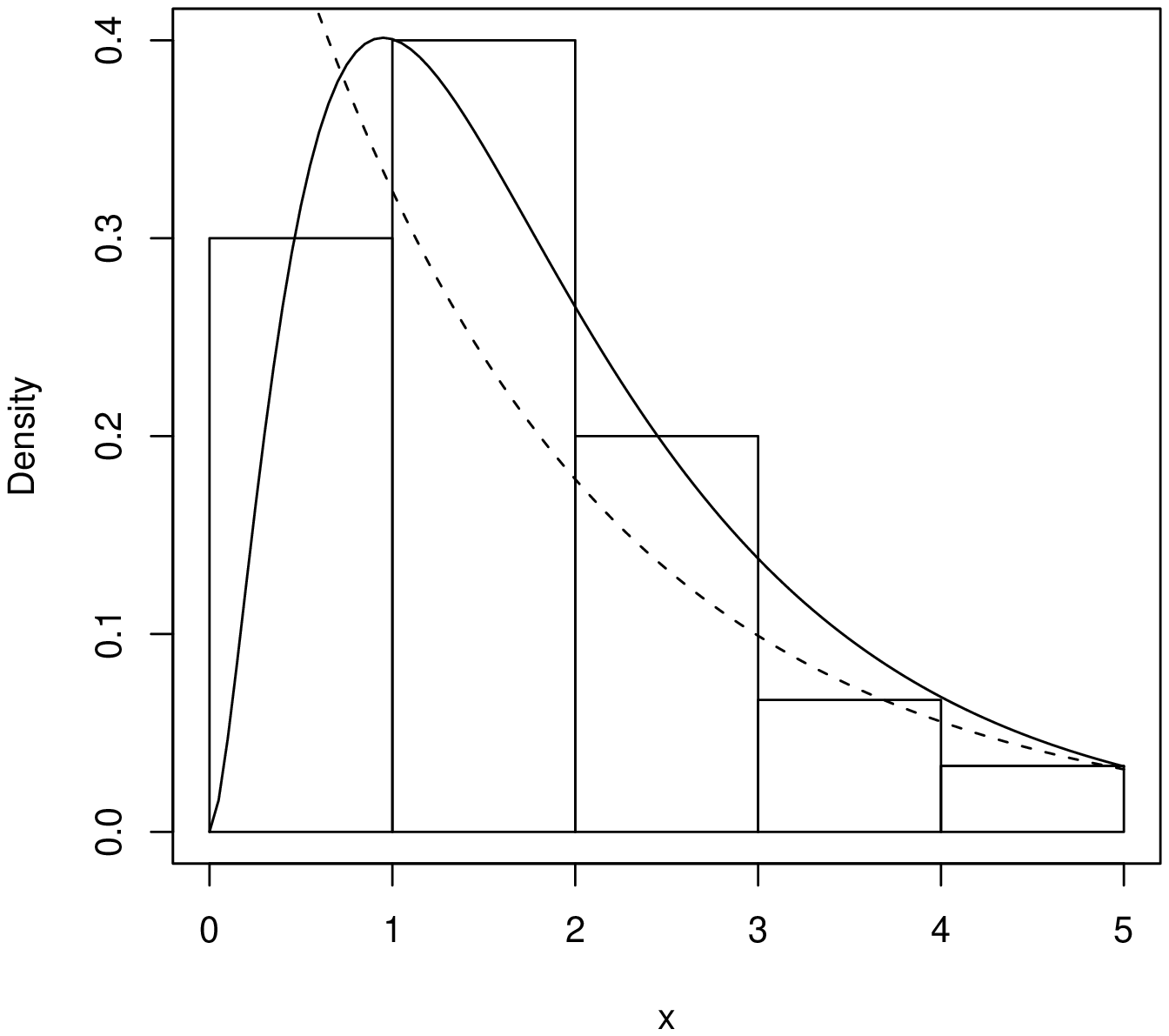}\includegraphics[width=0.50\textwidth]{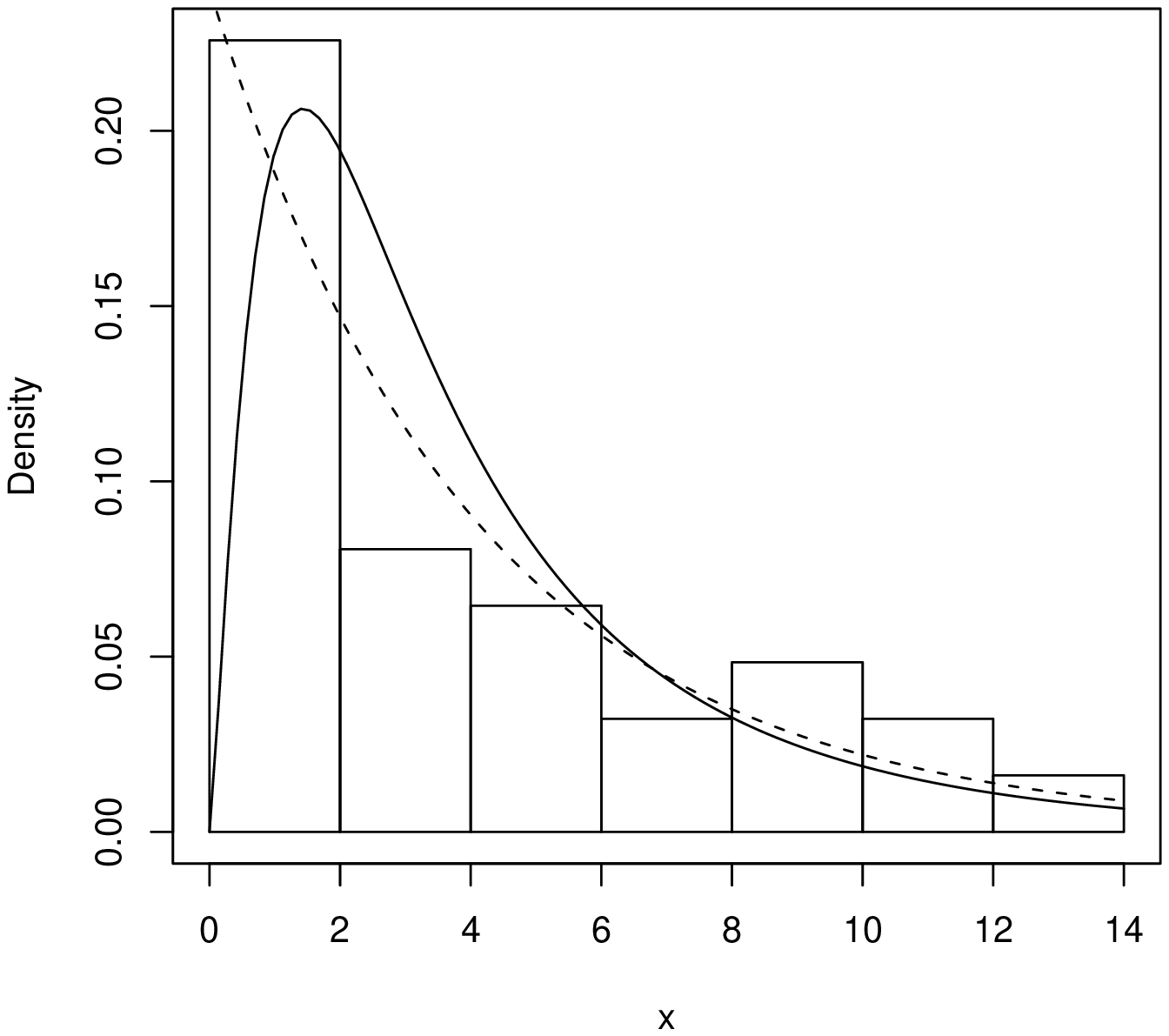}
	\caption{Fitted densities for the first (left) and second (right) data sets.}
\label{fig:figura3}
\end{figure}

We shall now wish test the null hypothesis $H_0: \mathrm{EP}$ against $H_1: \mathrm{GEP}$, i.e., $H_0: \alpha=1$ versus $H_1:\alpha\neq1$. The values of the $LR$ test statistic for the first and second data sets are 12.1412 ($p$-value: 4.9$\times 10^{-4}$) and 4.5636 ($p$-value: 3.2$\times10^{-2}$), respectively. Therefore, we reject the null hypothesis in both cases in favor of the GEP distribution at the usual significance levels. The plots of the fitted EP and GEP densities given in Figure \ref{fig:figura3} (together with the respective histograms) for the two data sets show that the GEP model yields better fits than the EP distribution.

\section{Concluding remarks}\label{S:conclusions}

We have generalized the exponential-Poisson (EP) distribution by defining the generalized exponential-Poisson (GEP) distribution. We derived important properties of the new distribution and obtained closed-form expressions for its moments and for order statistics moments. Applications to two real data sets were presented and discussed. In both applications, the GEP fit was superior to that obtained using the EP model.

\end{document}